# Quantum communication with pure state

**Arindam Mitra**

Anushakti Abasan, Uttar-Phalguni-7, 1/AF, Salt lake,

Kolkata, West Bengal, 700064, India.

**Abstract:** It is believed that quantum communication is not possible with a pure ensemble of states because quantum entropy of pure state is zero. This is indeed possible due to geometric consequence of entanglement.

Quantum entropy is the measure of information content of quantum state [1,2]. Only for mixed state quantum entropy is non-zero. Therefore, one may conclude that same pure ensemble cannot be used to represent bit 0 and 1. We shall see that same pure ensemble can represent bit 0 and 1 due to entanglement.

Suppose there is stock of EPR pairs. EPR pairs can be arranged in two different ways to represent bit 0 and 1. Let the two non-overlapping arrangements be

$$S_1 = \{A, B, B, C, D, A, E, F, E, F, C, D, ......\}$$
$$S_0 = \{B, A, D, C, D, A, E, E, F, F, B, C, ......\}$$

Here the same pair of letters denote an EPR pair, say singlet state of spin 1/2 particles. So, EPR state can be described as,

$$|\Psi\rangle_{i,j} = \frac{1}{\sqrt{2}}\left(|\uparrow\rangle_i|\downarrow\rangle_j - |\downarrow\rangle_i|\uparrow\rangle_j\right)$$

where i and j ($i \neq j$) denote the position of a pair in the sequence and "↑" and "↓" are two opposite spin directions. $S_1$ and $S_0$ are non-overlapping sequences because not a single pair occupies the same position ij in $S_1$ and $S_0$. Suppose Alice and Bob share the sequence codes. By sending sequences Alice can send message to Bob. This is basically entanglement-based alternative quantum coding (AQC) [3].

Alice transmits $S_0$ or $S_1$ to Bob. After receiving the sequence Bob measures spin components of all the particles along z axis. As Bob knows the sequence codes he recovers EPR correlation corresponding to either $S_0$ or $S_1$. That is, Bob recovers bit value recovering EPR correlation.

EPR pair gives any one of the two sets of data, "↑↓" and "↓↑" with equal probability. Non-EPR pair gives any one of the four sets of data, "↑↓", "↓↑", "↑↑" and "↓↓" with equal probability. Note that non-EPR pair also yields EPR data with probability 1/2.

It implies that by EPR correlation Bob cannot recover bit value with probability1. Increasing statistics in the two sequences bit value can be recovered with arbitrarily large probability. For non-overlapping sequences bit value can also be recovered with probability 1 in the following way.

First Bob makes a guess about the bit value. Then Bob sorts out a pair assuming non-EPR pair. Now he measures spin component of the pair. Conclusive identification of non-EPR pair is possible if Bob gets conclusive non-EPR data: "↑↑"or "↓↓". As Bob knows the sequence codes, conclusive recovery of bit value is possible with the conclusive identification of non-EPR pair. If the first chosen pair does not yield any conclusive non-EPR data, he guesses the other bit value and sorts out another pair assuming non-EPR pair and measures the spin component of the pair. Thus, alternately guessing two bit values, he sorts out pair assuming non-EPR pair and continues measurement until he gets conclusive non-EPR data. Here, Bob is alternately sorting out EPR pair and non-EPR pair since the sequences are non-overlapping. Therefore, Bob will quickly get conclusive non-EPR data to recover bit value with probability 1.

Although quantum entropy of this information source is zero, but Shannon entropy is log2. Another thing can be pointed out. A sequence of space-time points cannot completely specify a sequence of pure entangled states. Two or more arrangements of entangled states can be specified by the same sequence of space-time points. This is a geometric consequence of entanglement. These two things made quantum communication possible. This is a completely quantum coding which cannot have any classical counter-part. It can easily seen that two non-overlapping sequences of multi-particle entangled states can also be used for quantum communication.

The scheme also reveals that an unknown ensemble of known pure entangled states cannot be perfectly cloned. This has not been clearly revealed earlier. It may be recalled that standard no-cloning principle [4] tells that an unknown pure state cannot be cloned and von Neumann's no-cloning principle tells that an unknown mixture of known pure states cannot be cloned (in his famous equivalence of density matrix formalism, von Neumann essentially told [5] about the impossibility of cloning of quantum state, although cloning was not then a popular term). The coding reveals that space-time has direct bearing on the impossibility of cloning of quantum state. The scheme can be used to ensure [3] bit-by-bit security. Some other objectives can be fulfilled [6] within this entanglement-based AQC.

**References**
1. A. Peres, *Quantum theory: Concepts and Methods*, Kulwer Academic Publisher, Ch. 6.
2. M. A. Nielsen and I. L. Chuang, *Quantum information and computation*, Cambridge University Press, Ch. 11 & 12.
3. A. Mitra, http://xxx.lanl.gov/cs.IT/0501023.
4. W. K Wootters and W. J Zurek, *Nature* **299**, 802 (1981); D. Dieks, *Phy. Lett. A* **92**, 271 (1982).
5. v. Neumann, *Mathematical foundation of quantum mechanics* (Springer, Berlin, 1955) pp. 332
6. A. Mitra, http://xxx.lanl.gov/cs,CR/0502026; http://xxx.lanl.gov/cs,CR/0512007.